\documentclass [12pt]{article}
\def\maj#1{\ifmmode\mbox{\usefont{U}{msb}{m}{n}#1}\else{\usefont{U}{msb}{m}{n}#1}\fi}
\def\v#1{\mathbf{#1}}
\usepackage[dvips]{graphics,epsfig,graphicx}
\usepackage{color}

\begin{document}

\title{\textbf{Theory of spin precession\\ monitored by laser
pulse}}
\author{M. Combescot and O.
Betbeder-Matibet\\
\small{\textit{GPS, Universit\'e
Pierre et Marie Curie and Universit\'e Denis
Diderot, CNRS,}}\\
\small{\textit{Campus Boucicaut, 140 rue de
Lourmel, 75015 Paris, France}}}
\date{}
\maketitle

\begin{abstract}
We first predict the splitting of a spin
degenerate impurity level when this impurity
is irradiated by a circularly polarized laser
beam tuned in the transparency region of a
semiconductor. This splitting, which comes
from different exchange processes between the
impurity electron and the
virtual pairs coupled to the pump beam,
induces a spin precession around the laser
beam axis, which lasts as long as the pump
pulse. It can thus be used for ultrafast spin
manipulation. This effect, which has
similarities with the
exciton optical Stark effect we studied long
ago, is here derived using the concepts we
developed very recently to treat many-body
interactions between composite excitons and
which make the physics of this type of
effects quite transparent. They, in particular, allow to easily
extend this work to other experimental situations in which a
spin rotates under laser irradiation.
\end{abstract}

\vspace{2cm}

PACS.: 71.35.-y Excitons and related phenomena

\newpage

Long ago, Dani\`{e}le Hulin and her group [1]
discovered that, when a semiconductor is
irradiated by photons with energy too low to
create electron-hole pairs, the exciton line
blue-shifts. We have shown [2] that
this shift, which disappears when the pump
laser is turned off, comes
from interactions between the real
exciton created by the probe photon
and the virtual excitons coupled to
the pump beam [3].

In this communication, we predict an effect which
has similarities with this exciton optical
Stark shift: When an impurity is irradiated
by a pump beam tuned in the transparency
region of a semiconductor, its electronic
levels shift: The electron bound to the donor
interacts with the virtual electron-hole pairs
coupled to the pump beam, either by Coulomb
interaction, or by Pauli exclusion. If we
choose the pump polarization 
in such a way that the exchange processes
between the virtual pair and the up and down
electrons of the impurity are different, this
Pauli ``interaction'' splits the impurity
level. As a result, the spin of the
impurity electron precesses around the laser
beam axis, as long as the pump is turned on.
This effect can thus be used for ultrafast
spin manipulation, a subject of great technological interest in
the present days [4-10].

The impurity shift induced by a pump
beam is derived following a procedure
inspired from the one we used long ago to get
the exciton optical Stark shift [3]. However, to enlighten the
physics of this effect, we here calculate it using a
``commutation technique'' similar to the one
we recently developed for excitons
interacting with excitons [11] and which allows to identify
the two basic ingredients of the electron-virtual pair
interactions, namely a \emph{direct} Coulomb
scattering and a Pauli (or exchange) ``scattering'' ---
without any Coulomb contribution. The shift
results from the interplay
between the two, while the splitting only comes
from different carrier exchanges.

To make the physics which
controls the impurity level shift more
transparent, we, in the first part, assume that the
impurity electron and the electron of the virtual pairs have the
same spin. The spin degrees of freedom and the laser
polarization, of course crucial to get an impurity level
splitting, will be introduced in the second part.

We end this communication by reconsidering other
experimental conditions in which spins can rotate under laser
pulse, namely free electron in a quantum well [8,9] and
electrons trapped in quantum dots [5]. We explicitly show
how our present theory can be easily extended to these cases.

\vspace{0.5cm}

\noindent\textbf{Impurity level shift under laser irradiation}

Let us consider a semiconductor having a
ionized donor. Its Hamiltonian
reads $H'_{sc}=H_{sc}+W_I$, where
$H_{sc}=H_0+W_{sc}$ is the bare
semiconductor Hamiltonian, with $H_0=h_e+h_h$
and $W_{sc}=V_{ee}+V_{hh}+V_{eh}$, while
$W_I$ is the Coulomb interaction between the ionized donor and
the carriers. This interaction,
$\sum_n[-e^2/r_{e_n}+e^2/r_{h_n}]$, reads in
second quantization,
\begin{equation}
W_I=-\sum_{\v k,\v q}V_{\v q}\,a_{\v k+\v q}
^\dag\,a_{\v k}+\sum_{\v k,\v q}V_{\v q}\,
b_{\v k+\v q}^\dag\,b_{\v k}\ .
\end{equation}
$a_{\v k}^\dag$ and $b_{\v k}^\dag$ are the
creation operators for free electrons and holes, \emph{i}.\
\emph{e}., $(h_e-\epsilon_{\v k}^{(e)})a_{\v
k}^\dag|v\rangle=0$, while
$V_{\v q}=4\pi e^2/\mathcal{V}q^2$ or
$2\pi e^2/\mathcal{S}q$ are the Coulomb matrix elements
between free carriers, in 3D or 2D systems. In the presence of
the ionized donor,
the $H'_{sc}$ one-electron eigenstates, 
$(H'_{sc}-\epsilon_\mu)|f_\mu\rangle =0$, 
can be formally written as 
\begin{equation}
|f_\mu\rangle=a_\mu^\dag|v
\rangle=\sum_{\v k}\langle \v k|f_\mu\rangle
a_{\v k}^\dag|v\rangle\ . 
\end{equation}

If we now irradiate this system with pump photons
$(\omega_0,\v Q_0)$, the coupled matter-photon Hamiltonian
reads
$\mathcal{H}=H'_{sc}+H_{ph}+\mathcal{U}$,
where $H_{ph}=\omega_0c_0^\dag c_0$ is the bare
photon Hamiltonian and $\mathcal{U}=(U_0
^\dag c_0+h.c.)$ the semiconductor-photon
coupling. $U_0^\dag$,
which creates one electron-hole pair with
momentum $\v Q_0$, can be written as
$U_0^\dag=A^\ast\sum_{\v p}B_{\v p,\v Q_0}
^\dag$ where 
$B_{\v p,\v Q}^\dag=a_{\v p+\alpha_e\v
Q}^\dag
\,b_{-\v p+\alpha_h\v Q}^\dag$, with
$\alpha_e=1-\alpha_h=m_e/(m_e+m_h)$, 
is the creation operator for one free electron-hole pair [12]
with center of mass momentum $\v Q$ and
relative motion momentum
$\v p$. It is such that
$(H_0-E_g-E_{\v p,
\v Q})B_{\v p,\v Q}^\dag|v\rangle=0$, where
$E_g$ is the band gap, while
$E_{\v p,\v Q}=\hbar^2\v p^2/2m+\hbar^2\v Q^2
/2M$, with $m^{-1}=m_e^{-1}+m_h^{-1}$ and
$M=m_e+m_h$, is the $(\v p,\v Q)$ pair energy.

As for the exciton optical Stark effect [3], the
impurity level shift results from the
difference between the impurity level change
and the vacuum level change induced by the
pump beam. For
$\mathcal{U}=0$, the eigenstate with a ionized impurity
and $N_0$ photons is
$|v\rangle\otimes|N_0\rangle$, its energy
being $\mathcal{E}_0=N_0\omega_0$. At
lowest order in $\mathcal{U}$, this energy
becomes
\begin{equation}
\mathcal{E}'_0\simeq N_0\omega_0+N_0\langle
v|U_0\,
\frac{1}{\omega_0-H'_{sc}}\,U_0^\dag|v\rangle\
.
\end{equation}

In a similar way, for $\mathcal{U}=0$, 
the eigenstates with
one electron and $N_0$ photons are
$a_\mu ^\dag|v\rangle\otimes |N_0\rangle$,
their energy being
$\mathcal{E}_\mu=\epsilon_\mu +N_0\omega_0$,
while at lowest order in $\mathcal{U}$, they
read
\begin{equation}
\mathcal{E}'_\mu\simeq\epsilon_\mu+N_0\omega_0
+N_0\langle v|a_\mu\,U_0\,\frac{1}{\omega_0
+\epsilon_\mu-H'_{sc}}\,U_0^\dag\,a_\mu^\dag
|v\rangle\ ,
\end{equation}
The
impurity level shift induced by the pump
beam, $[\mathcal{E}'_\mu-\mathcal{E}'_0]-[
\mathcal{E}_\mu-\mathcal{E}_0]$, is thus
\begin{equation}
\Delta_\mu=N_0\langle v|U_0\left(a_\mu\,
\frac{1}{\omega_0+\epsilon_\mu-H'_{sc}}\,
a_\mu^\dag\,-\,\frac{1}{\omega_0-H'_{sc}}
\right)U_0^\dag|v\rangle\ .
\end{equation}
This quantity, linear in the pump intensity,
is formally similar to our expression of the
exciton optical Stark shift, with
$a_\mu^\dag$ and $\epsilon_\mu$ just
replacing the probe exciton creation operator
$B_t^\dag$ and energy $E_t$.

In order to calculate $\Delta_\mu$,
we introduce the Coulomb \emph{creation}
potential $V_\mu^\dag$, which, in this problem, is defined as
$[H'_{sc},a_\mu^\dag]=\epsilon_\mu a_\mu^\dag
+V_\mu^\dag$. It precisely reads
\begin{equation}
V_\mu^\dag=\sum_{\v k}\langle \v
k|f_\mu\rangle\sum_{\v q}V_{\v q}\,a_{\v k+\v
q}^\dag\sum_{\v k'}\left(a_{\v k'-\v q}^\dag\,
a_{\v k'}-b_{\v k'-\v q}^\dag\,
b_{\v k'}\right)\ .
\end{equation}
From the formal definition of $V_\mu^\dag$, it is easy to check
that
\begin{equation}
\frac{1}{x-H'_{sc}}\,a_\mu^\dag=a_\mu^\dag\,
\frac{1}{x-H'_{sc}-\epsilon_\mu}+\frac{1}{x-
H'_{sc}}\,V_\mu^\dag\,\frac{1}{x-H'_{sc}-
\epsilon_\mu}\ ,
\end{equation}
which is valid for any scalar
$x$. This allows to split $\Delta_\mu$ as
\begin{equation}
\Delta_\mu=N_0\,\left[\frac{1}{2}(\alpha_\mu+\alpha
_\mu^\ast)+\frac{1}{2}(\beta_\mu+\beta_\mu^\ast)
+\gamma_\mu\right]\ .
\end{equation}
Note that we have done a similar splitting in the
case of the exciton optical Stark shift [3]. 

As explicitly shown below, $\alpha_\mu$, given by
\begin{equation}
\alpha_\mu=\langle v|U_0\,a_\mu^\dag a_\mu\,
|\psi_0\rangle\ ,
\end{equation}
in which we have set $|\psi_0\rangle= 
(H'_{sc}-\omega_0)^{-1}U_0^\dag|v\rangle$,
comes from ``Pauli  interaction''
between the impurity electron and the virtual
pair. On the opposite, 
$\beta_\mu$ and $\gamma_\mu$, given by
\begin{eqnarray}
\beta_\mu&=&\langle\psi_0|a_\mu\,V_\mu^\dag|
\psi_0\rangle\nonumber\\
\gamma_\mu&=&
\langle\psi_0|V_\mu\,(\omega_0+\epsilon_\mu
-H_{sc}')^{-1}\,V_\mu^\dag|\psi_0\rangle\ ,
\end{eqnarray}
contain one or two $V_\mu^\dag$ operators, so that they come
from Coulomb interaction between
the impurity electron and the virtual pair. 

In the following, it will be convenient to
develop $|\psi_0\rangle$ on free pair states, according to
\begin{equation}
|\psi_0\rangle=A^{\ast}\sum_{\v
p,\v Q}
G(\v p,\v Q)\,B_{\v p,\v
Q}^\dag|v\rangle\ ,
\end{equation}
\begin{equation}
G(\v p,\v Q)=\sum_{\v p'}\langle v|B_{\v p,\v
Q}\,\frac{1}{H'_{sc}-\omega_0}\,B_{\v p',\v
Q_0}^\dag|v\rangle\ .
\end{equation}

\vspace{0.5cm}

\noindent\textbf{``Commutation technique'' for
a free pair interacting with an impurity electron}

$\alpha_\mu$ and
$\beta_\mu$ are easy to write in terms of the two 
``scatterings''
controling the physics of this problem, namely $\Lambda$ and 
$\Xi^\mathrm{dir}$, which appear in a
``commutation technique''
inspired from the one we recently developed
to treat many-body effects between composite
excitons [11]. From
\begin{equation}
\left[a_{\mu'}a_\mu^\dag\ ,B_{\v p,\v Q}^\dag
\right]=-\sum_{\v p',\v Q'}
\Lambda_{\mu'\v p'\v Q';\mu\v p\v Q}\,B_{\v
p',\v Q'}^\dag\ ,
\end{equation}
one of these two ``scatterings'', which is dimensionless, is
found to be
\begin{eqnarray}
\Lambda_{\mu'\v p'\v Q';\mu\v p\v Q}&=&\langle
f_{\mu'}|\v p+\alpha_e\v Q\rangle\,\langle
\v p'+\alpha_e \v Q'|f_\mu\rangle\,\delta_{
-\v p'+\alpha_h\v Q',-\v p+\alpha_h\v Q}
\nonumber\\
 &\equiv&\int
d\v r_e\,d\v r_{e'}\,d\v r_h\,
\langle f_{\mu'}|\v r_e\rangle\,\langle\v
p',\v Q'|\v r_{e'},\v r_h\rangle\,\langle
\v r_e,\v r_h|\v p,\v Q\rangle\,\langle
\v r_{e'}|f_\mu\rangle\ .
\end{eqnarray}
It corresponds to a bare electron exchange
between the impurity level $\mu$ and the
pair $(\v p,\v Q)$, which
transforms them into an ``out'' impurity level
$\mu'$ and an ``out'' pair $(\v
p',\v Q')$. Note that this $\Lambda$ scattering is Coulomb
free. From eq.\ (13), we can show that
\begin{equation}
\langle v|B_{\v p',\v
Q'}\,a_{\mu'}\,a_\mu^\dag\,B_{\v p,\v Q}^\dag
|v\rangle=\delta_{\mu',\mu}\,\delta_{\v p',\v
p}\,\delta_{\v Q',\v Q}-\Lambda_{\mu'\v p'\v
Q';\mu\v p\v Q}\ .
\end{equation}

The second scattering, defined through
\begin{equation}
\left[V_\mu^\dag,B_{\v p,\v Q}^\dag\right]=
\sum_{\mu',\v p',\v Q'}\Xi_{\mu'\v p'\v Q';
\mu\v p\v
Q}^\mathrm{dir}\,a_{\mu'}^\dag\,B_{\v p',\v
Q'} ^\dag\ ,
\end{equation}
is found to be
\begin{eqnarray}
\Xi_{\mu'\v p'\v Q';\mu\v p\v
Q}^\mathrm{dir}=V_{\v Q-\v Q'}\sum_{\v k,\v
k'}\langle f_{\mu'}|\v k'\rangle\,\langle \v
k|f_\mu\rangle\,
\delta_{\v k'+\v Q',\v k+\v Q}\left[
\delta_{-\v p'+\alpha_h\v Q',-\v p+\alpha_h\v
Q}-\delta_{\v p'+\alpha_e\v Q',\v
p+\alpha_e\v Q}\right]\nonumber\\
\equiv\int d\v r_e\,d\v r_{e'}\,d\v
r_h\,
\langle f_{\mu'}|\v r_{e'}\rangle\,\langle\v
p',\v Q'|\v r_e,\v
r_h\rangle[v_{e'e}-v_{e'h}]\,\langle
\v r_e,\v r_h|\v p,\v Q\rangle\,\langle
\v r_{e'}|f_\mu\rangle\ ,\hspace{0.4cm}
\end{eqnarray}
where $v_{ij}=e^2/|\v r_i-\v r_j|$. It corresponds
to \emph{direct} Coulomb interactions between
the impurity electron and the pair, \emph{without}
any carrier exchange. 

\vspace{0.5cm}

\noindent\textbf{Calculation of the impurity level shift}

\noindent (i) \textit{Pure Pauli term}

Equations (9,11,15) allow to write
$\alpha_\mu$ as
\begin{equation}
\alpha_\mu=|A|^2\sum_{\v p',\v p,\v Q}
\Lambda_{\mu\v p'\v Q_0;\mu\v p\v Q}\,
G(\v p,\v Q)\ .
\end{equation}
Equation (18) makes clear that this part of the shift is linked
to ``Pauli interaction'', \emph{i}.\ \emph{e}., 
exchange between the impurity electron and the virtual pairs. 
By noting that, at large detuning
$\Omega=E_g-\omega_0$, $G(\v
p,\v Q)$ tends to $\delta_{\v Q,\v
Q_0}/\Omega$, we can extract this limit from
$\alpha_\mu$ to write it as
$\alpha_\mu=|A|^2(1+\eta)/\Omega$, where, due to eqs.\ (12,14),
$\eta$ is precisely given by
\begin{equation}
\eta=\sum_{\v p',\v p,\v Q}\langle
\v p-\alpha_h\v Q+\v Q_0|f_\mu\rangle
\langle f_\mu|\v p+\alpha_e\v Q\rangle
\langle v|B_{\v p,\v
Q}\left[\frac{E_g-\omega_0}{H'_{sc}-\omega_0}
-1\right]B_{\v p',\v Q_0}^\dag|v\rangle\ .
\end{equation}
For large $\Omega$, the bracket of
eq.\ (19) tends to zero, so that $\alpha_\mu$
does reduce to
$|A|^2/\Omega$. This limit has to be compared to the one of
the similar Pauli term $\alpha$ in the optical Stark shift,
namely
$2 |A|^2/\Omega$. The link between these two limits can be
physically understood by noting that, in the case of the
exciton shift, a virtual pump pair can exchange both, its
electron and its hole, with the probe exciton, while here, it
can only exchange its electron with the impurity level: The
numerical prefactor of the large detuning leading term just
results from one carrier exchange instead of two.

In order to calculate the next order term $\eta$, we use 
\begin{equation}
\frac{1}{\omega_0-H'_{sc}}=
\frac{1}{\omega_0-H_0}+\frac{1}{\omega_0-H_0}
(W_I+W_{sc})\frac{1}{\omega_0-H_0}+\cdots\ ,
\end{equation}
which follows from $H'_{sc}=
H_0 +W_I+W_{sc}$.
The first term of eq.\ (20) leads to
replace $H'_{sc}$ by $H_0$ in eq.\ (19). As
$Q_0\ll 1/a_X$, while for bound states
$|\langle\v k|f_\mu\rangle|^2\simeq 0$ for
$k\gg 1/a_X$, we find that the contribution
of this term to $\eta$ is of the order of
$R_X/\Omega$, where
$R_X=\hbar^2/2ma_X^2$. A similar $R_X/\Omega$ behavior is
found for each of the two terms of $W_I$. On the opposite,
the $W_{sc}$ term of eq.\ (20), which
corresponds to Coulomb interaction
\emph{inside} the virtual pair (see fig.\ 1b), becomes singular
for large momentum transfers. It leads to
\begin{equation}
\eta\simeq\Omega\sum_{\v p,\v
p'}\frac{|\langle\v p +\alpha_e\v Q_0|
f_\mu\rangle|^2\,V_{\v p'-\v p}}{(\Omega+E_{\v
p,\v Q_0})(\Omega+E_{\v p',\v Q_0})}
\simeq\sum_{\v p'}\frac{V_{\v p'}}{\Omega+
\hbar^2\v p'^2/2m}=\tilde{\alpha}\,\sqrt{
\frac{R_X}{\Omega}}\ ,
\end{equation}
with $\tilde{\alpha}=2$ for 3D and
$\tilde{\alpha}=\pi$ for 2D; so that we end
with
\begin{equation}
\alpha_\mu=\frac{|A|^2}{\Omega}\left[1+
\tilde{\alpha}\sqrt{\frac{R_X}{\Omega}}+
O\left(\frac{R_X}{\Omega}\right)\right]\ .
\end{equation}
Note that, as $R_X\propto e^4$,
$\sqrt{R_X/\Omega}$ is in fact the dimensionless
parameter associated to a Coulomb expansion.

\noindent (ii) \textit{First order Coulomb term between the 
impurity electron and the virtual pairs}

We now turn to $\beta_\mu$.
Using eqs.\ (10,11,15,16), it reads
\begin{equation}
\beta_\mu=|A|^2\sum_{\v p',\v Q',\v p,\v Q}
G^\ast(\v p',\v Q')\left[\Xi_{\mu\v p'\v Q';\mu\v
p\v Q}^\mathrm{dir}-\Xi_{\mu\v p'\v Q';\mu\v
p\v Q}^\mathrm{in}\right]G(\v p,\v Q)\ ,
\end{equation}
where $\Xi_{\mu'\v p'\v Q';\mu\v
p\v Q}^\mathrm{in}$ is the sum
over $(\mu'',\v p'',\v Q'')$ of
$\Lambda_{\mu'\v p'\v Q';\mu''\v p''\v Q''}\,
\Xi_{\mu''\v p''\v Q'';\mu\v
p\v Q}^\mathrm{dir}$. Being made of a direct
Coulomb process \emph{between} the 
impurity electron and the pair, followed by an electron
exchange (see fig.\ 1c), $\Xi^\mathrm{in}$ is
actually an exchange Coulomb scattering.

To get the $\beta_\mu$ lowest order term in
$\sqrt{R_X/\Omega}$, \emph{i}.\ \emph{e}.,
in Coulomb interaction, we can replace
$H'_{sc}$ by its free carrier  expression
$H_0$, \emph{i}.\ \emph{e}., 
$G(\v p,\v Q)$ by $\delta_{\v Q,\v
Q_0}/(\Omega+E_{\v p,\v Q})$. The two terms of
$\Xi^\mathrm{dir}$ being then
equal, we are left with the exchange term,
which gives
\begin{eqnarray}
\beta_\mu&\simeq&|A|^2\sum_{\v k,\v p}|\langle
\v k|f_\mu\rangle|^2\frac{V_{\v p+\alpha_e\v
Q_0-\v k}}{\Omega+E_{\v p,\v Q_0}}\left[
-\frac{1}{\Omega+E_{\v p,\v Q_0}}+
\frac{1}{\Omega+E_{\v k-\alpha_e\v Q_0,\v
Q_0}}\right]\nonumber
\\ &\simeq&\frac{|A|^2}{\Omega}\left[\frac{
\tilde{\alpha}}{2}\sqrt{\frac{R_X}{\Omega}}
+O\left(\frac{R_X}{\Omega}\right)\right]\ .
\end{eqnarray}

\noindent (iii) \textit{Correlation term}

The last contribution $\gamma_\mu$ 
contains two Coulomb interactions between the impurity electron
and the free pair,
\emph{i}.\
\emph{e}., two ($e^2\propto \sqrt{R_X}$)
factors. In the large detuning limit, it thus behaves 
as $R_X/\Omega$ at least (other $e^2$ factors possibly appearing
if we expand $(\omega_0+\epsilon_\mu-H_{sc}')^{-1}$ according to
eq.\ (7)). Consequently, in this large detuning limit,
$\gamma_\mu$ is negligible in front of
$\alpha_\mu$ and $\beta_\mu$. On the opposite, the $\gamma_\mu$
contribution is the one possibly leading to resonances in the
impurity level shift. Indeed, if we look at eq.\ (10), we see
that $\gamma_\mu$ contains
$(\omega_0+\epsilon_\mu-H_{sc}')^{-1}$ acting on two electrons
plus one hole. The corresponding $H_{sc}'$ eigenstates being
the excitons bound to an impurity, we can inject the
closure relation for these states in front of this $H_{sc}'$
dependent operator.
$\gamma_\mu$ then shows poles at
$\omega_0=E_g+\hat{\epsilon}_\mu$, where the
$\hat{\epsilon}_\mu$'s are the energies of these excitons
bound on impurity.

This leads us to conclude that, at large
detuning, the impurity level shift
$|A|^2N_0/\Omega$ is entirely
controlled by electron exchange between
the impurity and the virtual pairs coupled
to the pump beam, without 
any Coulomb contribution (see fig.\ 1a). The
next order term, which is
$\sqrt{R_X}/
\Omega$ smaller, is also due to an electron
exchange but contains, in addition, one Coulomb
interaction, either inside the virtual pairs as in
$\alpha_\mu$ (see fig.\ 1b), or between
these virtual pairs and the impurity electron
as in
$\beta_\mu$ (see fig.\ 1c). On the opposite, possible 
resonances at the bound exciton energies can be found in the
correlation term $\gamma_\mu$, which, at large detuning, gives a
negligible contribution.

\vspace{0.5cm}

\noindent\textbf{Impurity level splitting}

Let us now see how the pump polarization and
the spin degrees of freedom affect these
results.

The semiconductor-photon coupling now reads
\begin{equation}
U_0^\dag=\sum_{\v p,s,m}A_{s,m}^\ast\,B_{\v
p,\v Q_0;s,m}^\dag\ ,
\end{equation}
where $B_{\v p,\v Q;s,m}^\dag=a_{\v
p+\alpha_e\v Q,s}^\dag b_{-\v p+\alpha_h\v
Q,m} ^\dag$ creates a pair with an electron
spin $s=\pm1/2$ and a hole momentum
$m=(\pm3/2,
\pm1/2)$ for bulk materials while $m=(\pm3/2)$
only for quantum wells.
The $A_{s,m}$'s depend on photon
polarization. For bulk materials, their
non-zero values are
$A_{\mp1/2,\pm3/2}=A_{\pm}$ and
$A_{\pm1/2,\pm1/2}=-A_{\pm}/\sqrt{3}$, while, for quantum wells,
these $A_{\pm1/2,\pm1/2}$'s are zero .
In the case of a circularly polarized beam
$\sigma_{\pm}$, the $A_{\pm}$'s are such that
$A_{\pm}=A$ and
$A_{\mp}=0$,
while for a linear beam along $x$ (resp.\
$y$), they are
$A_+=A_-=A/\sqrt{2}$ (resp.\
$A_+=-A_-=A/\sqrt{2}$).

In addition to these complexities in the semiconductor-photon
interaction, we have also to take into account the fact that the
impurity levels are now degenerate, the up and down spins
having the same energy --- in the absence of pump beam.
Consequently, it is now necessary to use degenerate
perturbation theory to get the impurity level
change induced by the laser beam. 
It is possible to show that
this change is obtained from the
diagonalization of a $2\times 2$ matrix, its
eigenvalues being
\begin{equation}
\mathcal{E}'_\mu=\epsilon_\mu+N_0\omega_0
+\frac{N_0}{2}\left[d_{++}+d_{--}\pm\sqrt{
(d_{++}-d_{--})^2+4|d_{+-}|^2}\right]\ ,
\end{equation}
\begin{equation}
d_{\sigma '\sigma}=\langle v|U_0\,a_{\mu,\sigma
'}\,\frac{1}{\omega_0+\epsilon_\mu
-H'_{sc}}\,a_{\mu,\sigma}
^\dag\,U_0^\dag|v\rangle\ .
\end{equation}
By taking into account the vacuum level
change induced by the pump beam, still given by
eq.\ (3), we end with an impurity
level having an average shift equal to
$\hat{\Delta}=N_0(\hat{d}_{++}+\hat{d}_{--})/2$,
and a splitting given by
$\hat{\delta}=N_0(\sqrt{(\hat{d}_{++}-
\hat{d}_{--})^2+4|\hat{d}_{+-}|^2}$,
where
\begin{equation}
\hat{d}_{\sigma'\sigma}=d_{\sigma'\sigma}-
\delta_{\sigma',\sigma}
\langle v|U_0(\omega_0-H'_{sc})^{-1}U_0^\dag
|v\rangle\ .
\end{equation}
Note that eq.\ (28) is a generalization of eq.\ (5), in the
presence of spin degrees of freedom.

To get these
$\hat{d}_{\sigma'\sigma}$, we use a
commutation technique similar to the one
without spin. In the presence of spins, the three scatterings
$\Xi^\mathrm{dir}$, $\Xi^\mathrm{in}$ and  $\Lambda$ are
now the product of an orbital part, which
is the one without spin, and a
spin part. Due to spin
conservation in Coulomb and exchange processes, this spin part
is just
$\delta_{\sigma',\sigma}\delta_{s',s}
\delta_{m',m}$ for the
direct scattering $\Xi^\mathrm{dir}$ (see
fig.\ (1d), and 
$\delta_{\sigma',s}\delta_{s',\sigma}
\delta_{m',m}$ for the exchange scatterings
$\Xi^\mathrm{in}$ and $\Lambda$ (see fig.\ 1e
).

It is then easy to show that, again, the
large detuning leading term of 
$\hat{d}_{\sigma'\sigma}$ is entirely
controlled by electron exchange
between the impurity and the virtual
pairs coupled to the pump, the next order term  having just
one additional Coulomb interaction either inside the
pair or between the pair and the impurity
electron. The two
first terms of
$N_0\hat{d}_{\sigma'\sigma}$ correspond
to the two first terms of 
$\Delta_\mu$ as obtained previously in eqs.\ (18) and (23), with
$|A|^2$ just replaced by
\begin{equation}
\pi_{\sigma'\sigma}=\sum_mA_{\sigma',m}^\ast\,
A_{\sigma,m}\ .
\end{equation}

We can then note that $\pi_{+-}=0$, since for a
given
$m$, there is only one $\sigma$ which makes
$A_{\sigma,m}\neq 0$, while $\pi_{\pm\pm}$
is equal to
$|A_{\mp}|^2+|A_{\pm}|^2/3$ for bulk
samples, and $|A_{\mp}|^2$ for quantum
wells. This shows that, when the pump beam is linear, 
$|A_+|=|A_-|$ so that $\pi_{++}=\pi_{--}$:
The impurity level has a blue shift equal to
$\Delta_\mu/2$ for quantum wells, and
$2\Delta_\mu/3$ for bulk materials, but no
splitting. On the opposite, for circular
beams, $A_+A_-=0$, so that
the impurity level splits: One impurity level
blue shifts of an amount $\Delta_\mu$, while
the other is unchanged for
quantum wells, or shifted by $\Delta_\mu/3$
for bulk materials. The splitting
$\hat{\delta}$ is then either $\Delta_\mu$ or
$2\Delta_\mu/3$.

\vspace{0.5cm}

\noindent\textbf{Spin precession of an impurity electron induced
by a pump beam}

Let us take $|\phi_0\rangle=(\cos\theta\,a_
{\mu+}^\dag+\sin\theta\,a_{\mu-}^\dag)|v
\rangle$ as initial impurity state. If we
turn on a circularly polarized pump beam
which propagates along $z$, the up and down
spins are shifted differently, due to their
different electron exchanges with the virtual pairs, so
that $|\phi_0\rangle$ becomes
\begin{equation}
|\phi_t\rangle=(\cos\theta\,a_{\mu+}^\dag
+e^{i\hat{\delta}t/\hbar}\,\sin\theta\,a_{\mu
-}^\dag)|v\rangle\ ,
\end{equation}
within a phase factor, $\hat{\delta}$ being the shift between
the $(\pm 1/2)$ impurity electrons calculated previously. The
projections of
$|\phi_t\rangle$ over $(+1/2)$ and $(-1/2)$ staying unchanged,
the spin of the impurity electron thus precesses
around the $z$ axis with a period
$T=2\pi\hbar/\hat{\delta}$. Since
$\hat{\delta}$ is of the order of
$\Delta_\mu$ --- which is just the
exciton optical Stark shift, within a factor
$1/2$, in the large detuning limit --- , we thus expect a
precession period of the order of 1psec within the experimental
conditions giving an exciton optical Stark
shift of the order of 1mev. We can note that this
period is far shorter than the spin
relaxation time, which is of the order of
1nsec.

\vspace{0.5cm}

\noindent \textbf{Extension of the theory to other spin
precessions} 

Let us end this communication by considering two cases in which
spin precession induced by laser beams has been described.

\noindent (i) \textit{Free electron in a quantum well} [8,9]

This case can be readily deduced from the above results by
setting the Coulomb potential between the carriers and the
ionized impurity $W_I$ equal to zero. This leads to replace
$a_\mu^\dag$ by $a_{\v k_0}^\dag$, $\epsilon_\mu$ by 
$\epsilon_{\v k_0}^{(e)}$ and
$|f_\mu\rangle$ by
$|\v k_0\rangle$, with $\langle\v k|\v k_0\rangle=\delta_{\v
k,\v k_0}$, in the formal expression of the shift $\Delta_\mu$
as well as in its $\alpha_\mu$, $\beta_\mu$ and $\gamma_\mu$
contributions. We have shown that, in the large detuning limit,
the two first terms of the shift are controlled by an electron
exchange, with possibly one Coulomb interaction inside the
virtual pairs or between these pairs and the impurity electron,
the Coulomb interaction with the ionized donor entering at
the next order,
$R_X/\Omega$, only. This shows that the shift and splitting of
the impurity electron and the ones of a free electron are
thus just the same for these two large detuning terms, provided
that
$\epsilon_{\v k_0}^{(e)}\ll \Omega$, for eq.\ (21) to be valid.
On the opposite, the possible resonances coming from the
$\gamma_\mu$ contribution differ. They are now controlled by
the two electron-one hole eigenstates, \emph{i}.\ \emph{e}.,
the trions, while, in the presence of impurity, they are
controlled by excitons bound to the impurity.
As the coupling between photon and trion is in fact
extremely weak in the large sample limit [13], the weights of
these resonances are expected to be rather small.

\noindent (ii) \textit{Electron in a quantum dot} [5]

The spin precession of an electron trapped in a quantum dot
can also be deduced from the above theory. The one-body
electron Hamiltonian $h_e$ has just to now include the
dot confinement. Instead of $a_{\v k}^\dag$, the
creation operator for a Coulomb free electron reads
$a_n^\dag$, with 
$(h_e-\epsilon_n^{(e)})\,a_n^\dag|v\rangle=0$. These
eigenstates \emph{a priori} include bound states as well as
extended states, if the barrier height is finite.

If we now consider one electron trapped in the dot ground state
$a_{n_0}^\dag|v\rangle$, its shift $\Delta_{n_0}$ is given by
eq.\ (5), with $a_\mu^\dag$ and $\epsilon_\mu$ replaced by 
$a_{n_0}^\dag$ and $\epsilon_{n_0}^{(e)}$, while the dot-photon
coupling has now to be written as
$U_0^\dag=\sum_{n,m}A_{nm}^\ast a_n^\dag b_m^\dag$. This shift
can be calculated using a ``commutation technique'' formally
similar, the Coulomb \emph{creation} potential now
reading
\begin{equation}
V_{n_0}^\dag=\sum_na_n^\dag\sum_{m',m}\left[V_{ee}\left(^n_{m'}\
^{n_0}_m\right)a_{m'}^\dag a_m+V_{eh}\left(^n_{m'}\
^{n_0}_m\right)b_{m'}^\dag b_m\right]\ ,
\end{equation}
where $V_{ee}\left(^{n'}_{m'}\ ^n_m\right)$ and 
$V_{eh}\left(^{n'}_{m'}\ ^n_m\right)$ are the Coulomb matrix
elements between dot states $(n,m)$ and $(n',m')$.

The calculation of $\Delta_{nÑ0}$ which is performed in a
quite similar way, shows that, at large detuning, the shift is
again controlled by electron exchange, its leading term now
reading 
$N_0\Omega^{-1}\sum_m|A_{n_0m}|^2$, while resonant
contributions in the correlation term $\gamma_\mu$ must appear
at the eigenenergies of a ``trion'' in the dot. 

\vspace{0.5cm}

\noindent\textbf{Conclusion}

We have shown that, due to carrier exchanges
between the impurity and the virtual pairs
coupled to a pump beam tuned in the
transparency region of a semiconductor, the
up and down electronic levels of an impurity blue shift. The
degenerate levels of this impurity can also
split if the pump beam is circularly
polarized, due to differences in these carrier exchanges. This
splitting induces a spin precession around the laser beam axis,
which lasts as long as the pulse. It can thus be used to
manipulate spins. We have also shown how the present theory can
be extended to other spin precessions induced by laser beam,
such as the one of free electrons in a quantum well or the one
of electrons trapped in a quantum dot.

\vspace{0.5cm}

We wish to thank J. Tribollet for inducing
this work and C. Mora for stimulating discussions.

\vspace{1cm}

\noindent REFERENCES

\vspace{0.5cm}

\noindent
[1] A. Mysyrowicz, D. Hulin, A. Antonetti, A. Migus, W.T.
Masselink, H. Morko\c{c}, \emph{Phys.\ Rev.\ Lett.\ }
\textbf{56}, 2748 (1986).

\noindent
[2] For a review, see M. Combescot, \emph{Phys.\ Rep.\ }
\textbf{221}, 168 (1992).

\noindent
[3] M. Combescot, R. Combescot, \emph{Phys.\ Rev.\ Lett.\ }
\textbf{61}, 117 (1988); \emph{Phys.\ Rev.\ } B \textbf{40},
3191 (1989). M. Combescot, \emph{Phys.\ Rev.\ } B \textbf{41},
3517 (1990).

\noindent
[4] D. Loss, D.P. Di Vincenzo, \emph{Phys.\ Rev.\ } A
\textbf{57}, 120 (1998).

\noindent
[5] A. Imamoglu, D.D. Awschalom, G. Burkard, P.D. Di
Vincenzo, D. Loss, M. Sherwin, A. Small, \emph{Phys.\ Rev.\
Lett.\ } 
\textbf{83}, 4204 (1999).

\noindent
[6] P. Chen, C. Piermarocchi, L.J. Sham, \emph{Phys.\ Rev.\
Lett.\ } \textbf{87}, 67401 (2001).

\noindent
[7] C. Piermarocchi, P. Chen, L.J. Sham, D.G. Steel,
\emph{Phys.\ Rev.\ Lett.\ } \textbf{89}, 167402 (2002).

\noindent
[8] J.A. Gupta, R. Knobel, M. Samarth, D.D. Awschalom,
\emph{Science} \textbf{292}, 2459 (2002).

\noindent
[9] J. Bao, A.V. Bragas, J.K. Furdyna, R. Merlin, \emph{Nat.\
Mater.\ } \textbf{2}, 175 (2003).

\noindent
[10] P. Chen, C Piermarocchi, L.J. Sham, D. Gammon, D.G. Steel,
\emph{Phys.\ Rev.\ } B \textbf{69}, 75320 (2004).

\noindent
[11] M. Combescot, O. Betbeder-Matibet, \emph{Europhys.\ 
Lett.\ } \textbf{58}, 87 (2002). O. Betbeder-Matibet, M.
Combescot, \emph{Eur.\ Phys.\ J.\ } B \textbf{27}, 505 (2002).

\noindent
[12] While the semiconductor-photon interaction can also be
written in terms of excitons, it is in fact more convenient 
to write it in terms
of free electron-hole pairs for problems dealing with large
detunings.

\noindent
[13] M. Combescot, J. Tribollet, \emph{Solid State Com.\ }
\textbf{128}, 273 (2003).

\newpage

\begin{figure}
\centerline{ \scalebox{0.8}{\includegraphics{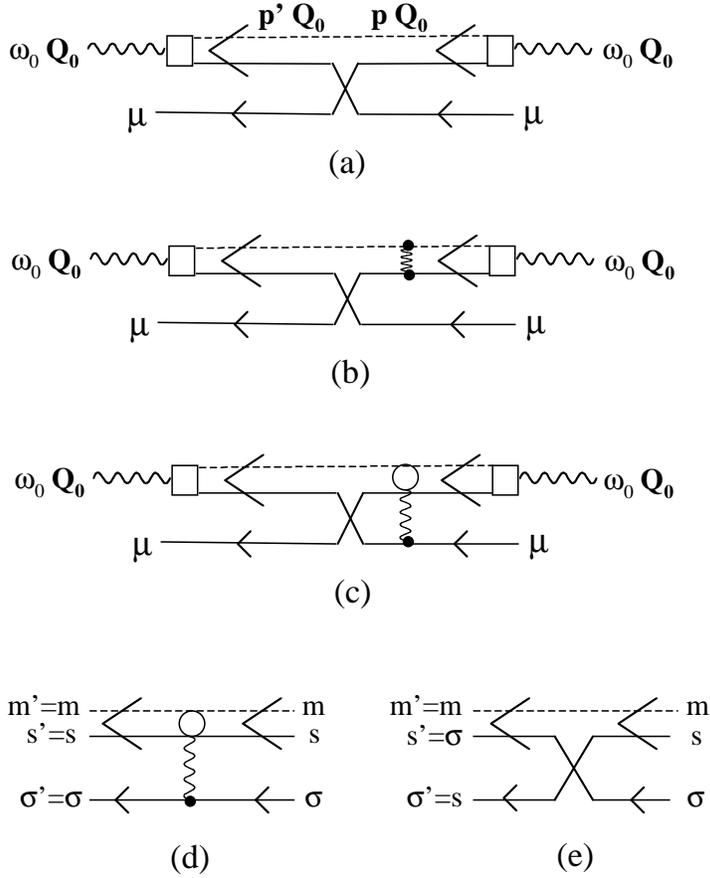}}}
\caption{(a): A photon $(\omega_0\v Q_0)$ creates a virtual
electron-hole pair $(\v p,\v Q_0)$. This pair exchanges its
electron with the electron of an impurity (in a $\mu$ state)
and finally recombines to give back the $(\omega_0\v Q_0)$
photon. This process is the dominant one in the impurity level
shift at large detuning. (b,c): The large detuning next order
term contains one Coulomb interaction either inside the
virtual pair (b) or between this pair and the impurity
electron (c). (d): The direct Coulomb scattering
$\Xi^\mathrm{dir}$ of the ``commutation technique'' for a free
pair interacting with an impurity electron: The ``in'' and
``out'' pairs are made with the same electron. (e): Exchange or
Pauli ``scattering'' $\Lambda$ of this commutation technique.
Note that this scattering exists in the absence of any Coulomb
process.}
\end{figure}

\end{document}